\documentstyle[11pt,newpasp,twoside]{article}
\def\edcomment#1{\iffalse\marginpar{\raggedright\sl#1\/}\else\relax\fi}
\reversemarginpar

\input{psfig.sty}

\begin{document}
\title{Radio Observations of the Magnetic Fields in Galaxies}
 \author{M. Krause}
\affil{Max-Planck-Institut f\"ur Radioastronomie, Auf dem H\"ugel 68,
       D-53121 Bonn, Germany}

\begin{abstract}
After a short introduction on how we get information of the magnetic
fields from radio observations I discuss the results concerning the
magnetic field structure in galaxies: Large-scale regular magnetic
field pattern of spiral structure exist in grand-design spirals,
flocculent and even irregular galaxies. The regular field in spirals is
aligned along the optical spiral arms but strongest in the interarm
region, sometimes forming `magnetic arms'. The strongest total field is
found in the optical arms, but mainly irregular. The large-scale
regular field is best explained  by some kind of dynamo action. Only a
few galaxies show a dominant axisymmetric field pattern, most
field structures seem to be a superposition of different dynamo modes
or rather reveal more local effects related to density waves, bars or
shocks. Observations of edge-on galaxies show that the magnetic fields
are mainly parallel to the disk except in some galaxies with strong
star formation and strong galactic winds as e.g. NGC~4631.
\end{abstract}

\section{Introduction or What do we get from Radio Observations?}

Radio observations of the continuum emission are best suitable to
study the magnetic fields in galaxies. The total intensity of the
synchrotron emission gives the strength of the total magnetic field.
The linearly polarized intensity reveals the strength and the structure
of the resolved regular field in the plane of the sky.  However, the
observed polarization vectors suffer Faraday rotation and
depolarization (i.e. a decrease of the degree of linear polarization
when compared to the intrinsic one) on the way from the radiation's
origin to us. Correction for Faraday rotation is possible with
observations at two or better more wavelengths by determining the
rotation measure RM (being proportional to $\int n_{\rm e} B_{\parallel}
dl$). The rotation measure itself is a measure of the magnetic field
strength  parallel to the line of sight, whereas its sign gives the
direction of this magnetic field component. The field strength of both
components, parallel and perpendicular to the line of sight, together
with the information of the intrinsic polarization vectors enables us
in principle to perform a `tomography' of the magnetic field.

\section{Faraday Rotation and Depolarization Effects}

Figure~1 gives an example of observations of M51 at 4 different
wavelengths, all smoothed to the same linear resolution of
$75\arcsec$ HPBW. The vectors are rotated by $90\deg$ but not corrected
for Faraday rotation. The figure illustrates nicely the different
effects of Faraday rotation and depolarization effects depending on
the observing wavelength: the observed vectors at $\lambda$2.8~cm and
$\lambda$6~cm are mainly parallel to the optical spiral arms as
expected in spiral galaxies (see below), Faraday rotation is small at
centimeter wavelengths. However, the pattern looks very different at
$\lambda$18/20~cm where Faraday rotation is expected to be strong.
Further, we see a region in the northeastern part of M51 with complete
depolarization.

\begin{figure}[htb]
\centerline{\psfig{figure=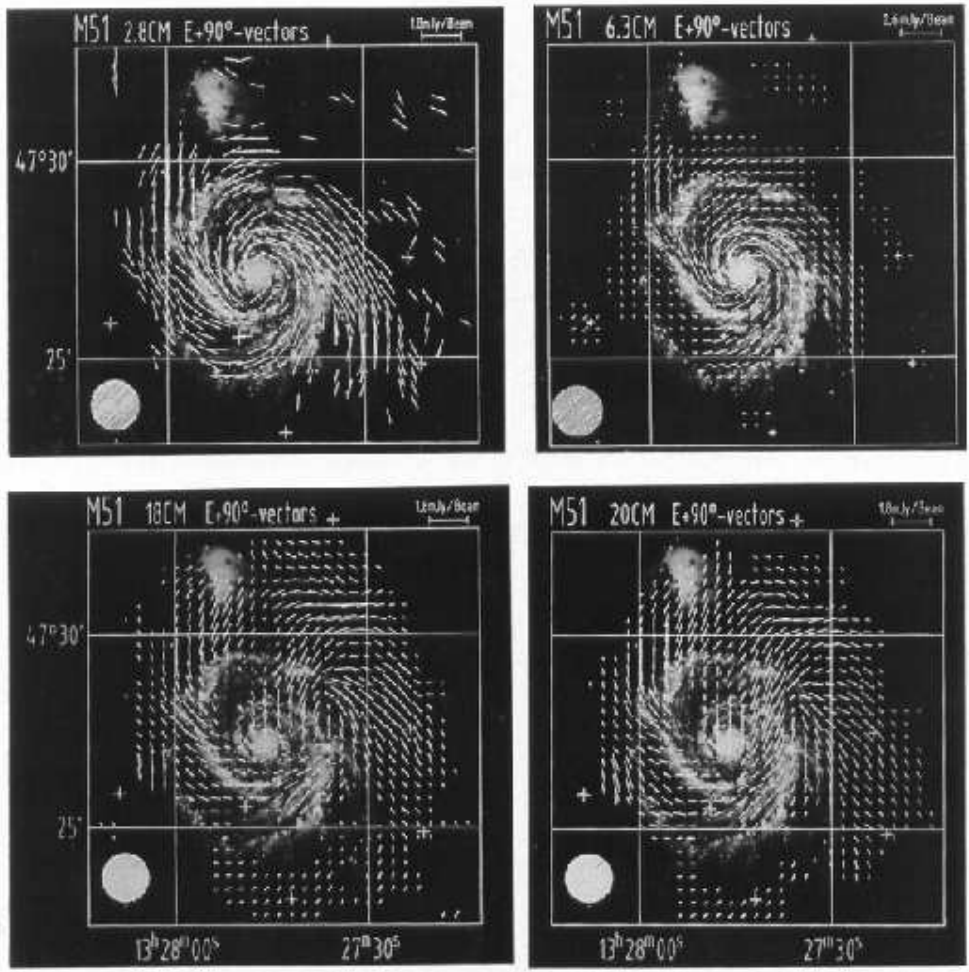,width=13truecm,%
        bbllx=45pt,bblly=135pt,bburx=567pt,bbury=657pt}}
\caption{Maps of the E-vectors rotated by $90\deg$ of M51 observed at
$\lambda\lambda$2.8~cm, 6.2~cm, 18.0~cm, and 20.5~cm. The length of the
vectors is proportional to the polarized intensity. They are shown
superimposed onto an optical picture (Lick Observatory).}
\end{figure}

After substraction of the thermal fraction of the emission we
distinguish between beam-dependent and wavelength-dependent
depolarization. The difference in depolarization at different
wavelengths in maps with the same linear resolution should be
purely wavelength dependent where two different wavelength-dependent
depolarization effects are important to consider: the differential
Faraday rotation and Faraday dispersion as despcribed by  Burn (1966)
and Sokoloff et al. (1998). The latter effect is due to turbulent
magnetic fields within the source and between the source and us,
whereas the Faraday rotation depends on the regular magnetic field
within the emitting source. The differential Faraday rotation has a
strong wavelength dependence as shown e.g. in Fig.~1 in
Sokoloff et al. (1998) leading to a complete depolarization at
$\lambda$20~cm already at a $\rm RM \approx 40\, rad/m^2$, with again
decreasing depolarization for higher RMs. Such an effect has first been
detected in small isolated areas in M51 (Horellou et al. 1992). At
$\lambda$6~cm the depolarization is much smaller, increasing smoothly
to zero at $\rm RM \approx 400\, rad/m^2$ (the first zero point is at
$\rm RM=\pi/(2 \cdot \lambda^2)$).

Hence, the galaxies may not be transparent in linear polarization at
decimeter wavelengths so that we may observe just an upper layer of the
whole disk. At centimeter wavelengths we do not expect complete
depolarization even in galaxies viewed edge-on, i.e. centimeter
wavelengths are best suitable to trace the magnetic field structure.

\section{Magnetic Field Strength and Structure}

The total magnetic field strength in a galaxy can be estimated from
the nonthermal radio emission under the assumption of equipartition
between the energies of the magnetic field and the relativistic
particles (the so called {\bf energy equipartition}).  The degree of
linear polarization and some assumptions of the geometry of the
magnetic field give the strength of the magnetic field that has a
uniform direction within the beam size. The estimates are based on the
formulae given by e.g. Pacholczyk (1970) and Segalowitz et al. (1976),
and are described e.g. by Krause et al. (1984) and Beck (1991).

\subsection{... in Spiral Galaxies}

The magnetic field has been found to be mainly parallel to the
galactic disk and to show a large spiral pattern similar to that of the
optical spiral arms. The total magnetic field strength is generally
highest at the positions of the optical spiral arms, whereas the highest
regular fields are found {\em offset} of the optical arms and in the
interarm region.

The mean equipartition value for the total magnetic field strength for
a sample of 74 spiral galaxies observed by Niklas (1995) is on average
$8\, \mu$G with a standard deviation of $3\,\mu$G. It can however
reach values of about $20\,\mu$G {\em within} spiral arm regions as
e.g. in NGC~6946 (Beck 1991). Strongly interacting galaxies or galaxies
with a strong central radio emission tend to have generally stronger
total magnetic fields. The strengths of the {\em regular} fields are
typically  1--5~$\mu$G in the {\em interarm} regions in nearby spirals
but reach locally $13\,\mu$G in NGC~6946 (Beck \& Hoernes 1996).

\subsection{Regular Magnetic Fields and Dynamo Action}

The large-scale magnetic field is generally thought to be amplified and
maintained by  the action of a large-scale dynamo. According the mean
field dynamo theory (e.g. Ruzmaikin et al. 1988; Wielebinski \&
F.~Krause 1993; Beck et al. 1996; Lesch \& Chiba 1997) the structure
of the large-scale field is also given by the dynamo action. It is
generally of spiral shape with different azimuthal field directions and
symmetries. The mode that can be excited most easily is the
axisymmetric mode ({\bf ASS}) followed by higher modes as the
bisymmetric ({\bf BSS}), etc. The  field configurations can be either
symmetric (quadrupole type) or asymmetric (dipole type) with respect
to the galactic plane. According to the dynamo theory the pitch angle
of the magnetic field spiral is determined by the dynamo numbers, not
by the pitch angle of the gaseous spiral arms.

ASS and BSS field configurations can be distinguished observationally
by analyzing the rotation measures or -- more sophisticated -- by
analyzing directly the observed polarization vectors at different
wavelengths (as has been described e.g. in Sokoloff et al. (1992) and
Berkhuijsen et al. (1997)). It has been found that only M31 and IC342
show clear ASS fields (Beck 1982; Krause et al. 1989a), whereas many
other galaxies seem to have a superposition of different modes.

A special case is M81 as it has a {\em dominating} BSS field field
(Krause et al. 1998b; Sokoloff et al. 1992). The dominance of the BSS
field structure requires additional physical mechanism to be invoked
that can occur only in rare cases. For M81 a three-dimensional,
nonlinear dynamo model has been developed including the disturbed
velocity field due to the encounter with its companion NGC~3077
(Moss et al. 1993) or alternatively, parametric resonance with the
spiral density wave as has been proposed by Chiba \& Tosa (1990) and
investigated numerically by Moss (1996).

Most other spiral galaxies observed so far indicate a mixture of
magnetic modes. The analysis of the observations at all 4 frequencies
of M51 (Fig.~1) revealed even two different magnetic field
configurations for the disk and the halo resp. (Berkhuijsen et al.
1997): a halo with an axisymmetric field configuration parallel to the
disk with magnetic field lines pointing {\em inwards} and a
superposition of an axisymmetric and a bisymmetric field with about
equal weights in the disk. The magnetic field lines in the disk are
spirals generally directed {\em outwards} except in a few sectors in
the inner northwestern part of M51 as shown in Fig.~2.

\begin{figure}[htb]
\centerline{\psfig{figure=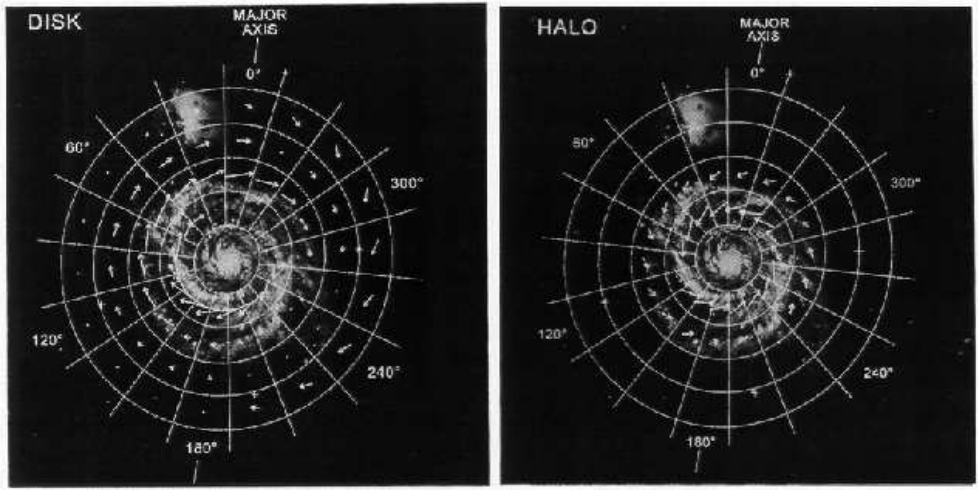,width=13truecm,%
        bbllx=-11pt,bblly=235pt,bburx=624pt,bbury=558pt}}
\caption{The vectors give the directions of the horizontal regular
magnetic field in the disk (left) and the halo (right) of M51 as fitted
to the observations at 4 different wavelengths in Fig.~1. They are shown
superimposed onto an optical picture (Lick Observatory). Note that the
magnetic field direction in the disk is opposite to that in the halo,
except in a few sectors in the inner northwestern part of the galaxy.}
\end{figure}

It is clear that such a global description of the regular magnetic
field cannot describe local effects that become more and more visible
with increasing linear resolution of the observations. Further, in
grand-design galaxies the magnetic field lines follow quite often the
dust lanes. In M51 e.g. one dust lane in the eastern part crosses the
optical spiral arm, so does the regular magnetic field (visible in
Fig.~1 at in the short wavelengths observations).

\subsection{Magnetic Arms}

The regular interarm field in M81 fills the whole interarm region.
Different to this some galaxies host so-called {\em magnetic arms}.
Long, highly polarized arms disconnected from the optical spiral arms
were first discovered in IC342 (Krause et al. 1989a; Krause 1993). Two
symmetric magnetic arms running right in the interarm region parallel
to the optical spiral arms were found in NGC~6946. Their width is less
than 1~kpc, hence they do not fill the whole interarm region. Magnetic
arms have also been found in NGC~2997 (Han et al. 1999) and M83 (Beck,
Ehle \& Sukumar in prep.).

Several models have been developed to explain these magnetic arms.
Slow MHD waves have been proposed by Fan \& Lou (1996) and Lou \& Fan
(1998) to explain the generation of magnetic arms shifted with respect
to the optical spiral arms. These MHD waves occur only in an almost
rigidly rotating disk. However, all galaxies with magnetic arms are
found to rotate differentially beyond 1--2~kpc from the center,
different to older measurements with lower angular resolution. Han et
al. (1999) found some correlation between the magnetic arms and
interarm gas features generated at the 4:1 resonances in numerical
models of Patsis et al. (1997).

In the framework of dynamo theory the generation of magnetic arms can be
described if one considers that the turbulent velocity of the
gas is higher in the optical arms (Moss 1998; Shukurov 1998), or that
turbulent diffusion is larger in the arms (Rohde et al. 1999). Both
effects reduce the dynamo number in the spiral arms when compared to the
interarm regions and hence allow to generate magnetic arms preferently
in the interarm region.

\subsection{... in Flocculent and Irregular Galaxies}

Regular magnetic fields have also been detected in flocculent galaxies
like M33 (Buczilowski \& Beck 1991) and NGC~4414 (Soida et al. 2002).
These are galaxies with a flocculent spiral structure without signs of
the action of density waves. The mean degree of polarization (corrected
for different angular resolution) is similar between flocculent and
grand-design galaxies (Knapik et al. 2000). As expected from classical
$\alpha - \Omega$ dynamo models the dynamo works well without the
assistance of density waves.

Even in a dwarf irregular galaxy with weak rotation and non-systematic
gas motion like NGC~4449 a large-scale (partly spiral) regular magnetic
field has been observed (Chyzy et al. 2000). The strength of the
regular field reaches $7\,\mu$G and that of the total field $14\,\mu$G,
which is high even in comparison with fields strengths of radio-bright
spirals. The absence of ordered differential rotation requires a
different kind of dynamo action in this galaxy. A fast field
amplification is predicted by a dynamo e.g. driven by magnetic buoyancy
and sheared Parker instabilities (e.g. Moss et al. 1999; Hanasz \&
Lesch 1998, 2000) or without any $\alpha$ effect at all (Blackman 1998).

\subsection{... in Barred Galaxies and Other Shocked Areas}

A sample of 20 barred galaxies has been observed extensively in total
power and linear polarization (Beck et al. 2002). They found that the
total radio emission (and hence the total magnetic field) is strongest
along the bar and correlates with the bar {\em length}. The regular
magnetic field is enhanced {\em upstream} of the shock fronts in the
bar. The upstream field lines are at large angle to the bar, but turn
sharply towards the bar about 1~kpc upstream from the dust lanes as
observed in NGC~1097 (Beck et al. 1999). According to these authors
similar effects have also been observed in NGC~1365, NGC~1672, and
NGC~7552.

Indications for a compression of the galactic magnetic field possibly
by gas tidally stripped during an interaction with the neighbouring
galaxy have been observed in NGC~3627 (Soida et al. 2001) where the
observed regular field apparently crosses the dust lanes at a large
angle in the east. Another example  for such a compression is the
wind-swept galaxy NGC~4254 (Soida et al. 1996).

\subsection{Edge-on Galaxies and Vertical Fields}

Several galaxies seen edge-on have been observed in radio continuum and
polarization. Most of them have regular magnetic fields that are
parallel to the galactic disks (Dumke et al. 1995), only a few are
found with large-scale vertical fields like M82 (Reuter et al. 1994),
NGC~4631 (Golla \& Hummel 1994), NGC~4666 (Dahlem et al. 1997), and
NGC~5775 (T\"ullmann et al. 2000).

The apparent disk thicknesses vary quite a lot among the galaxies with
plane-parallel field as well as their intensities does. Interferometer
observations of edge-on galaxies have to be combined with single-dish
observations in order to correct for the missing zero-spacings before
the scale heights of the emission perpendicular to the disk (in
z-direction) can be determined. We found that the emission in
z-direction can best be fitted with two exponential functions, whose
scale heights are about {\em equal} for all four galaxies with
plane-parallel fields that have been analyzed so far, namely
NGC~891, NGC~3628, NGC~4565, NGC~5907 (Dumke \& Krause 1998; Dumke et
al. 2000). The scale height for the thin disk is $\simeq 300$~pc and
that of the thick disk/halo is $\simeq 1.8$~kpc for these galaxies,
independent of the star-forming activity and interaction state.

\begin{figure}
\begin{minipage}[t]{6.4truecm}
\psfig{figure=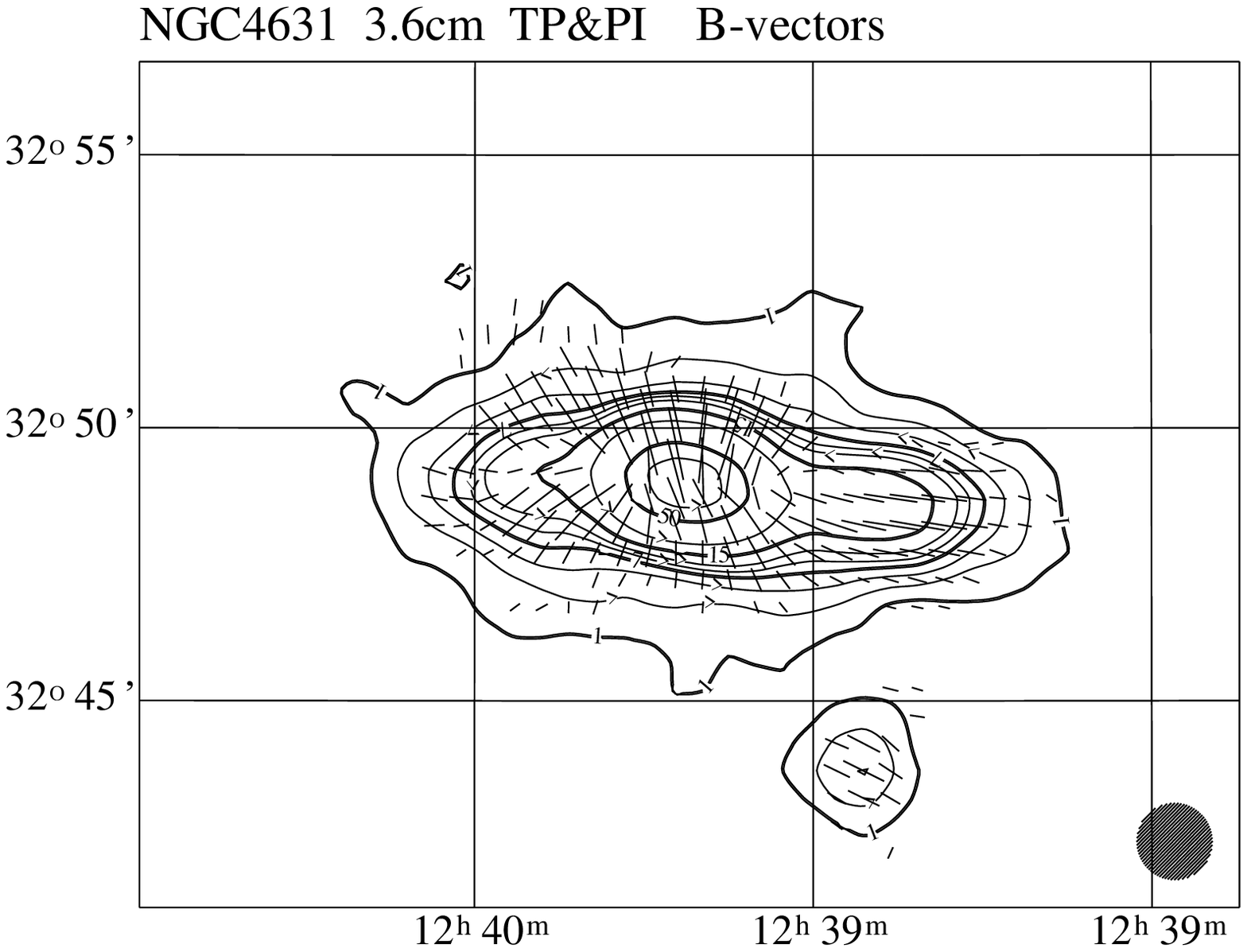,width=6.4truecm,%
        bbllx=60pt,bblly=138pt,bburx=516pt,bbury=486pt}
\caption{Contour map of NGC~4631 at $\lambda$3.6~cm as observed with
the Effelsberg 100-m telescope. The angular resolution is $85\arcsec$
HPBW. The vectors give the orientation of the intrinsic regular field
in the plane of the sky. Their lengths are proportional to the
polarized intensity.}
\end{minipage}\hfill
\begin{minipage}[t]{6.4truecm}
\psfig{figure=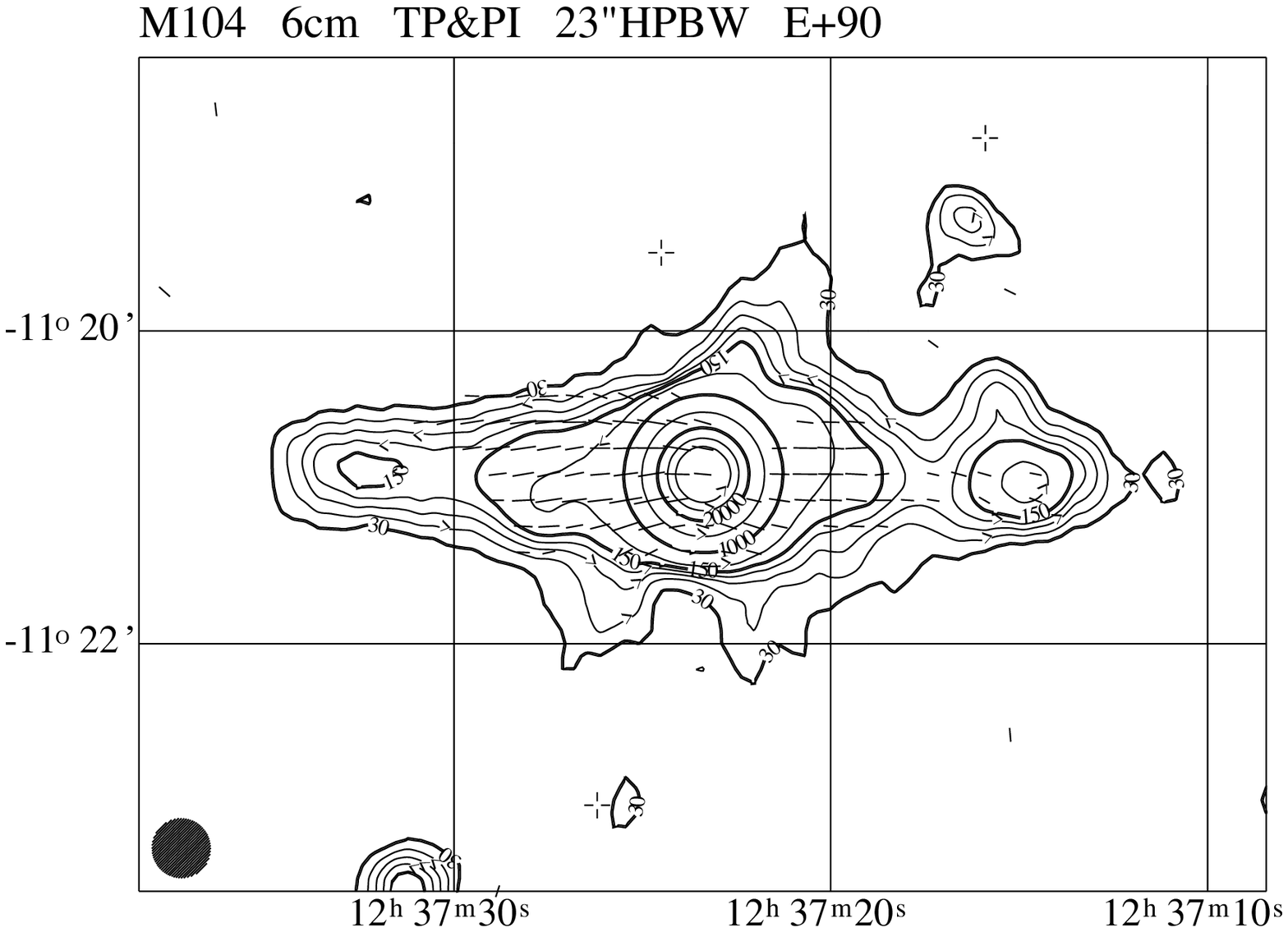,width=6.4truecm,%
        bbllx=66pt,bblly=277pt,bburx=573pt,bbury=644pt}
\caption{Contour map of M104 at $\lambda$6.2~cm as observed with the
VLA in D array. The angular resolution is $23\arcsec$ HPBW. The vectors
give the observed E-vectors rotated by $90\deg$, their lengths are
proportional to the polarized intensity.}
\end{minipage}
\end{figure}

Recent $\lambda$3.6~cm observations of NGC~4631 obtained with the
Effelsberg 100-m telescope are presented in Fig.~3. Faraday rotation
could be determined between this wavelength and $\lambda$6.2~cm
observations with the VLA and revealed $\rm -300\, rad/m^2 < RM < 300\,
rad/m^2$. The vectors shown in Fig.~3 give the intrinsic magnetic field
orientation. NGC~4631 has a large-scale vertical magnetic field in the
central 7~kpc. Outside this radius the field is plane parallel in the
western half but still has vertical field components in the
eastern half. The RM does not show a typical symmetric pattern as
expected from a dipole or quadrupole field. Hence we conclude that the
vertical field is rather wind-driven and related to the high star-forming
activity in this galaxy. The exponential scale heights for NGC~4631 are
about 50\% larger than those found for galaxies with plane-parallel
fields which may also be related to the galactic winds and
vertical fields.

Another edge-on galaxy is M104, the Sombrero galaxy, classified as an
Sa galaxy and known for its huge bulge. We observed M104 at
$\lambda$6.2~cm with the VLA in D array and detected that the observed
E+90\deg\  vectors are surprisingly regular and mainly parallel to the
galactic disk as shown in Fig.~4. Unfortunately, the observations
could not yet be corrected for Faraday rotation because we have no
observations at another wavelength. The regularity of the vectors at
$\lambda$6.2~cm may indicate that Faraday rotation is rather small.
Hence it seems that the regular magnetic field in M104 is mainly disk
parallel, also inside the central 6~kpc where the rotation curve is
still rising. The radio emission in z-direction can best be fitted by a
one-component Gaussian rather than an exponential function with a scale
height of $\simeq 3$~kpc. According to Combes (1991) a Gaussian
z-distribution is just expected for a thin disk inside a
self-gravitating mass distribution, i.e. the huge bulge in M104.

\end{document}